\input phyzzx
\input phyzzx.plus
\overfullrule=0pt
\def\sym{\mathop{\rm sym}\nolimits}
\REF\SUZ{H.~Suzuki, Nucl.\ Phys.\ B 585 (2000) 471.}
\REF\FUJ{T. Fujiwara, H. Suzuki and K. Wu, Nucl.\ Phys.\ B 569 (2000)
643.}
\REF\IGA{H.~Igarashi, K.~Okuyama and H.~Suzuki, in preparation.}
\Pubnum={IU-MSTP/45}
\titlepage
\title{Errata and Addenda to\break
``Anomaly Cancellation Condition in Lattice Gauge Theory''}
\author{Hiroshi Igarashi,\foot{igarashi@serra.sci.ibaraki.ac.jp}
Kiyoshi Okuyama\foot{okuyama@serra.sci.ibaraki.ac.jp}
and Hiroshi Suzuki\foot{hsuzuki@mito.ipc.ibaraki.ac.jp}}
\address{Department of Mathematical Sciences, Ibaraki University, Mito
310-8512, Japan}
\abstract{We correct some intermediate expressions and arguments in
Nucl.\ Phys.\ B 585 (2000) 471--513. The main results do not change.
We also mention some additional observations, including a constraint
on a coefficient of the possible nontrivial anomaly which was not
given in the paper.}
\bigskip\bigskip\bigskip
\noindent
PACS numbers: 11.15.Ha, 11.30.Rd\hfill\break
Keywords: chiral gauge theory, lattice gauge theory
\endpage
The main result of Ref.~[\SUZ] is the theorems stated in~Section~3
which determine the general structure of gauge anomalies in lattice
gauge theory. These theorems are based on local solutions to the
consistency condition in abelian theory with the ghost number
unity~$g=1$, Eq.~(6.24). We found that, although the formula~(6.24)
for~$g=1$ and thus the theorems in~Section~3 remain correct, some
intermediate expressions and arguments for general~$g$ were wrong.
Here we show how these must be corrected.

The field~$\widetilde\omega_\mu^{a_0[a_1\cdots a_g]}$ in~Eq.~(6.8) is
totally antisymmetric $\widetilde\omega_\mu^{a_0\cdots a_g}=%
\widetilde\omega_\mu^{[a_0\cdots a_g]}$ in nontrivial solutions as
shown in~Eq.~(6.12). The argument to show this through Eqs.~(6.10)
and~(6.11) is however wrong for general~$g$ and is corrected as
follows. We consider the first three lines of~Eq.~(6.8). By making use
of~$c^{a_0}(n+\widehat\mu)=c^{a_0}(n)+\delta_BA_\mu(n)$, one sees that
$$
\eqalign{
   &\biggl[
   A_\mu^{a_0}(n)c^{a_1}(n)\cdots c^{a_g}(n)
\cr
   &\quad
   -c^{a_0}(n+\widehat\mu)\sum_{i=1}^g{g\choose i}
   A_\mu^{a_1}(n)\delta_BA_\mu^{a_2}(n)\cdots\delta_BA_\mu^{a_i}(n)
   c^{a_{i+1}}(n)\cdots c^{a_g}(n)
\cr
   &\qquad
   -(\hbox{the totally antisymmetic part on $a_0$, $a_1$, $\cdots$,
   $a_g$})\biggr]\widetilde\omega_\mu^{a_0[a_1\cdots a_g]}(n)
\cr
   &=\delta_B\biggl[\biggl(
   -\sum_{i=1}^g{g!\over(i+1)!\,(g-i)!}\,A_\mu^{a_0}A_\mu^{a_1}
   \delta_BA_\mu^{a_2}\cdots\delta_BA_\mu^{a_i}
   c^{a_{i+1}}\cdots c^{a_g}\biggr)
   \widetilde\omega_\mu^{a_0[a_1\cdots a_g]}\biggr],
\cr
}
\eqn\one
$$
where the totally antisymmetric part of a quantity~$t^{a_0\cdots a_g}$
is defined
by~$\sum_\sigma\epsilon_\sigma t^{\sigma(a_0)\cdots\sigma(a_g)}
/(g+1)!$. Eq.~\one\ shows that only the totally antisymmetric part
of~$A_\mu^{a_0}c^{a_1}\cdots c^{a_g}-\cdots$ contributes to the
nontrivial part. As the result, we can assume that
$\widetilde\omega_\mu^{a_0[a_1\cdots a_g]}$ is totally antisymmetric
in nontrivial solutions, as shown in~Eq.~(6.12).

The coefficient in~Eq.~(6.13) must be chosen as
$$
   \widetilde\omega_\mu^{[a_0\cdots a_g]}(n)
   ={1\over3!}\varepsilon_{\mu\nu\rho\sigma}
   {(-1)^{g(g+1)/2}\over g+1}
   \Omega_{\nu\rho\sigma}^{[a_0\cdots a_g]}(n+\widehat\mu),
\eqn\two
$$
for the normalization of~Eq.~(6.14).

Eq.~(6.23) which shows the symmetry of the coefficients $B_2$
and~$B_0$ in~Eq.~(6.19) is wrong for general~$g$ and the derivation
through Eqs.~(6.20), (6.21) and~(6.22) is replaced as follows. We
first note that Eq.~(6.19) can be written as
$$
\eqalign{
   {\cal A}\,d^4x(d\theta)^g
   &\simeq\sum_n\Bigl[
   \sym(C^{a_1}\cdots C^{a_g}){\cal L}^{[a_1\cdots a_g]}\,d^4x
   +\sym(C^{a_1}\cdots C^{a_g})B_4^{[a_1\cdots a_g]}
\cr
   &\qquad\quad
   +\sym(C^{a_1}\cdots C^{a_g}F^b)B_2^{[a_1\cdots a_g]\,b}
   +\sym(C^{a_1}\cdots C^{a_g}F^bF^c)B_0^{[a_1\cdots a_g](bc)}
\cr
   &\qquad\quad
   +\sym(A^{a_0}C^{a_1}\cdots C^{a_g})B_3^{[a_0\cdots a_g]}
   +\sym(A^{a_0}C^{a_1}\cdots C^{a_g}F^b)B_1^{[a_0\cdots a_g]\,b}
   \Bigr].
\cr
}
\eqn\three
$$
Namely, the field strength 2-forms~$F^b$ can be put in the
symmetrization symbol in spite of the noncommutativity of differential
forms. If we substitute each coefficients in this expression by
totally antisymmetrized ones including one of indices for the field
strength 2-forms,
$B_2^{[a_1\cdots a_g]\,b}\to B_2^{[a_1\cdots a_gb]}$,
$B_0^{[a_1\cdots a_g](bc)}\to B_0^{[a_1\cdots a_gb]c}$ and
$B_1^{[a_0\cdots a_g]\,b}\to B_1^{[a_0\cdots a_gb]}$, we see that
$$
\eqalign{
   &\sum_n\sym(C^{a_1}\cdots C^{a_g}F^{b_1}F^{b_2}\cdots F^{b_r})
   B^{[a_1\cdots a_gb_1]\,b_2\cdots b_r}
\cr
   &=s\sum_n{g\over2}\sym(A^{b_1}A^{a_1}C^{a_2}\cdots C^{a_g}
   F^{b_2}\cdots F^{b_r})
   B^{[a_1\cdots a_gb_1]\,b_2\cdots b_r},
\cr
}
\eqn\four
$$
and
$$
\eqalign{
   &\sum_n
   \sym(A^{a_0}C^{a_1}\cdots C^{a_g}F^{b_1}F^{b_2}\cdots F^{b_r})
   B^{[a_0\cdots a_gb_1]\,b_2\cdots b_r}
\cr
   &=s\sum_n{g\over3!}\sym(A^{b_1}A^{a_0}A^{a_1}C^{a_2}\cdots C^{a_g}
   F^{b_2}\cdots F^{b_r})
   B^{[a_0\cdots a_gb_1]\,b_2\cdots b_r}.
\cr
}
\eqn\five
$$
As the result, the following antisymmetric parts of the coefficients
$B_2^{[a_1\cdots a_g]\,b}$, $B_0^{[a_1\cdots a_g](bc)}$
and~$B_1^{[a_0\cdots a_g]\,b}$ in~Eq.~\three\ can be set to zero
because they contribute only to BRS trivial parts:
$$
   B_2^{[a_1\cdots a_gb]}=0,\qquad
   B_0^{[a_1\cdots a_gb]c}=0,\qquad
   B_1^{[a_0\cdots a_gb]}=0.
\eqn\six
$$
The first two equations replace Eq.~(6.23) and the last one gives rise
to the constraint on~$B_1^{[a_0\cdots a_g]\,b}$ which was not given
in~Ref.~[\SUZ]. Actually, Eq.~\six\ is identical to the constraints
for corresponding coefficients in the continuum theory. For solutions
with the ghost number unity, $g=1$, the first two constraints
in~Eq.~\six\ are equivalent to~Eq.~(6.23). Therefore, Eq.~(6.24) which
is for~$g=1$ holds as it stands.

A quick way to see the equivalence of Eq.~\three\ and~Eq.~(6.19) is to
introduce the superspace derivative~$\widetilde d=d+s$ and the
superspace connection~$\widetilde A^a=A^a+C^a$. We see that
$\widetilde d\widetilde A^a=dA^a=F^a$ (the horizontality condition)
and~$\widetilde dF^a=0$. With this language, we can write, for
example,
$$
\eqalign{
   &\sym(A^{a_0}C^{a_1}\cdots C^{a_g}F^{b_1}\cdots F^{b_r})
   B^{[a_0\cdots a_g](b_1\cdots b_r)}
\cr
   &={1\over g+1}
   \sym(\widetilde A^{a_0}\widetilde A^{a_1}\cdots \widetilde A^{a_g}
   F^{b_1}\cdots F^{b_r})
   B^{[a_0\cdots a_g](b_1\cdots b_r)}\bigr|_{O(d\theta^g)}.
\cr
}
\eqn\seven
$$
Then we consider its difference to the combination
$$
   {1\over g+1}
   \sym(\widetilde A^{a_0}\widetilde A^{a_1}\cdots \widetilde A^{a_g})
   F^{b_1}\cdots F^{b_r}
   B^{[a_0\cdots a_g](b_1\cdots b_r)}\bigr|_{O(d\theta^g)}.
\eqn\eight
$$
An exchange of $\widetilde A^a$ and~$F^b$, according to the
noncommutative differential calculus~[\FUJ], produces the
commutator\foot{From this, one immediately sees that
$[F^a,F^b]=2\widetilde d\varphi_3^{ab}=2d\varphi_3^{ab}$.}
$$
   [\widetilde A^a,F^b]=\widetilde dY_2^{ab}+2\varphi_3^{ab},
\eqn\nine
$$
(see Eqs.~(5.47), (5.48) and~(5.52) of~Ref.~[\SUZ]) where $Y_2^{ab}$
and~$\varphi_3^{ab}$ do not contain~$d\theta$ and $\varphi_3^{ab}$
depends only on the field strength (see Eqs.~(5.49) and~(5.54)\foot{%
Eq.~(5.54) of~Ref.~[\SUZ] must be replaced by
$$
\eqalign{
   \varphi_3(n)&={1\over12}\bigl[
   F_{\alpha\beta}^a(n)F_{\beta\gamma}^b(n)
   +2F_{\alpha\beta}^a(n+\widehat\gamma)F_{\beta\gamma}^b(n)
\cr
   &\qquad\quad
   +2F_{\alpha\beta}^a(n)F_{\beta\gamma}^b(n+\widehat\alpha)
   +F_{\alpha\beta}^a(n+\widehat\gamma)
   F_{\beta\gamma}^b(n+\widehat\alpha)\bigr]
   dx_\alpha\,dx_\beta\,dx_\gamma B_0^{[ab]}.
\cr
}
\eqn\ten
$$
}
of~Ref.~[\SUZ]). Under the summation~$\sum_n$, one can do the
integration by parts with respect to~$\widetilde d$ up to BRS trivial
terms. After this integration by parts, $Y_2^{ab}$ in the commutator
does not contribute to~$O(d\theta^g)$-term in~Eqs.~\seven\ and~\eight,
because $Y_2^{ab}$ and~$\widetilde d\widetilde A^a=F^a$ do not
contain~$d\theta$ (recall that $\widetilde dF^b=0$). Namely, under the
summation~$\sum_n$, we can neglect~$Y_2^{ab}$ in the commutator, up to
BRS trivial terms. $\varphi_3^{ab}$ in the commutator on the other
hand cannot be neglected. However, it is easy to see that its
contribution can be absorbed into the first term of the right hand
side of~Eq.~\three\ up to BRS trivial terms, because $\varphi_3^{ab}$
depends only on the field strength. In this way, we see the
equivalence of Eq.~\three\ and~Eq.~(6.19). This superspace trick can
also be applied to derive Eqs.~\four\ and~\five.

The cumbersome proof of the covariant Poincar\'e lemma for~$G=U(1)^N$
in~Ref.~[\SUZ] was limited for 4- or lower-dimensional lattice. It is
however possible to give a simpler proof which works for arbitrary
dimensional lattices. This proof, a detailed study of nontrivial local
solutions to the consistency condition with an arbitrary ghost number
and its applications will be given elsewhere~[\IGA].

\refout
\bye